\documentstyle[jkas2]{article}

\runningauthor{Shang}
\runningtitle{Structure and Physical Conditions in MHD Jets}
\beginpage{1}
\endpage{7}

\begin{document}

\title{Structure and Physical Conditions in MHD Jets from Young Stars}

\author{Hsien Shang}

\address{Institute of Astronomy and Astrophysics, Academia Sinica, Taipei 106,
TAIWAN \\
Center for Astrophysics, 60 Garden St., MS-42, Cambridge, MA 02138, USA\\
{\it E-mail:shang@asiaa.sinica.edu.tw}}

\address{\normalsize{\it (Received 09/15, 2001; Accepted ???. ??,2001)}}

\abstract{
We have constructed the foundations to a series of theoretical diagnostic
methods to probe the jet phenomenon in young stars as observed at various
optical forbidden lines.  We calculate and model in a self-consistent manner
the physical and radiative processes which arise within an inner disk-wind
driven magnetocentrifugally from the circumstellar accretion disk of a young
sun-like star.  Comparing with real data taken at high angular resolution, our
approach will provide the basis of systematic diagnostics for jets and their
related young stellar objects, to attest the emission mechanisms of such
phenomena.  This work can help bring first-principle theoretical predictions to
confront actual multi-wavelength observations, and will bridge the link between
many very sophiscated numerical simulations and observational data.
Analysis methods discussed here are immediately applicable to new
high-resolution data obtained with HST and Adaptic Optics.}


\maketitle

\section{Introduction}
Herbig-Haro objects (Herbig 1950,Haro 1950) are small nebulous objects
that trace jet-like, collimated structures, with characteristic emission spectra
of hydrogen and optical forbidden lines of [O~I], [N~II], and [S~II], in the
red wavelengths.  Mechanisms that produce these phenomena play necessary roles
in the making of young stars.

Whether jets alone can also drive the often associated molecular outflows
has been long debated (Reipurth \& Bally 2001).  The morphology and the
distribution of mass and momenta of these outflows argue for wide-angle winds
pushing the ejecta and sweeping up materials at wide solid angles (Shu et al
1991).  X-winds, by construction, are wide-angle winds, launched
magneto-centrifugally near the innermost edge of the circumstellar disks of
YSOs (Shu et al 1994), that expand and quickly fill space to ultimately
collimate at large distances.  Shang et al. (1998) demonstrated the density
structure and kinematic information for an x-wind based on a semi-analytic
method (Shang 1998), by computing emissions from [S~II]$\lambda$6716 and
[O~I]$\lambda$6300 lines with uniform ionization conditions throughout the
smooth flow without identifying the sources of excitation.  The synthetic
images for the forbidden lines show strongly cylindrically stratified
structures in the center, and suggested the visual appearance of jets is an
{\it optical illusion} out of an intrinsically very divergent flow (Shu et
al 1995).

It is very compelling to learn whether conditions arising self-consistently in
flows driven by magnetically interacting young star-disk systems can reproduce
the many observed features known about the dynamics, morphology, and excitation
conditions of jets and Herbig-Haro objects.  To identify a definitive set of
diagnostic tools to help probe the highly collimated jet emissions from the
theoretical aspects, is the primary task.  This also serves as the first test
to bring a well-developed MHD model based on semi-analytic approaches into
stringiest confrontation with observations.  The fundamental concepts and
approaches can be generalized with many multiwavelength observations, and
transferred to other theoretical MHD wind/jet models, such as the many
interesting numerical simulations discussed in this volume.

\section{Physics Conditions and Thermal Structure of Jets}

We opened new investigations of thermal structures in the MHD winds from young
stars, following an earlier attempt of Ruden, Glassgold, \& Shu (1990, RGS),
for a cold spherical stellar wind.  In RGS, they attempted a spherically
symmetric model, with density, velocity and Lorentz forces enough to accelerate
the neutrals to escape velocity.  They found that the dominant source of
heat for the gas is ambipolar diffusion associated with the magnetic
acceleration, and adiabatic expansion of the wind is the most powerful cooling
mechanism.  However, ambipolar diffusion failed to heat the wind beyond ${\sim 10
\;{\bf R_\ast}}$, and the plasma could be at most lightly ionized (${< 10^{-4}}$).
In the environment of star-disk interaction, many new processes have been
discovered and identified since RGS (e.g. Shu et al 1997).  In Shang et al
(2001), we incorporated the processes into the self-consistent x-wind flow and
made it a diagnostic package fore the jet phenomena.

New ingredients: X-rays, as observed to accompany essentially all YSOs (e.g.,
Feigelson \& Montmerle 1999), UV photons from accretion funnels onto stars, and
ambipolar diffusion, are among the most important radiative and physical
processes intrinsic to the MHD flows that ionize and heat the gas.
Many significant updates of physics in X-rays, hydrogen reactions and
momentum transfer cross sections and coefficients of ambipolar diffusion,
go into streamline-by-streamline computations of electron fraction, temperature
and abundances of important species.  The flow spans up to 8000 AU, right from
where the wind is launched.  This is the first {\it a priori} whole scale
computation of thermal structures of MHD winds by self-consistent flow
structure and processes that drive the winds.
\begin {figure}[t]
\vskip -0.5cm
\centerline{\epsfysize=8cm\epsfbox{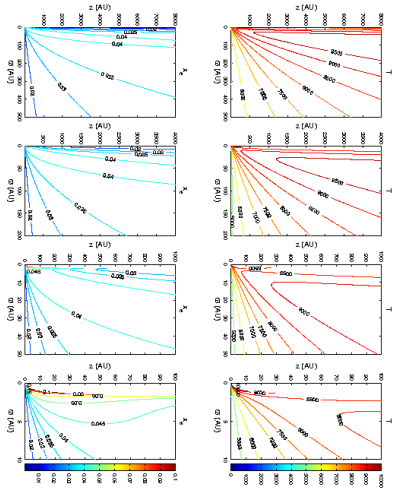}}
\caption{
Temperature (right) and ionization (left) contours
in the $\varpi-z$ plane for a fiducial case of an early but slightly revealed
YSO.  The units for the spatial scales are AU.
}
\end{figure}

\begin {figure}[t]
\vskip -0.5cm
\centerline{\epsfysize=8cm\epsfbox{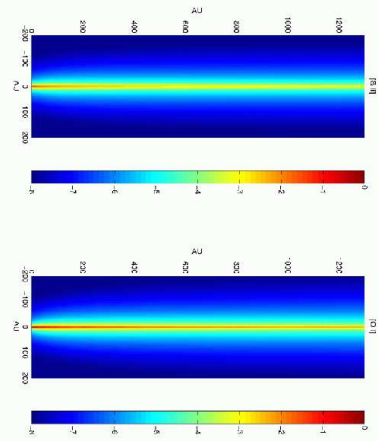}}
\caption{
Synthetic images
of the [S\,II] $\lambda$6731 (left) and [O\,I] $\lambda$6300 (right) brightness
for the same model as in Figure 1. The $\log_{10}$ of the integrated intensity
is plotted in units of erg s$^{-1}$ cm$^{-2}$ ster$^{-1}$.
}
\end{figure}


Studies with the approach by Shang et al (2001) concluded that x-rays are
capable of ionizing the base of the x-wind, and help maintain the ionization
level at large distances, which are balanced by slow radiative recombination.
This finding explains the profile of ionization in jets, which Bacciotti et al
(1995) inferred phenomenologically from ground-based data with forbidden line
ratios, but whose ionization source was not identified.  Better analysis of
Bacciotti \& Eisl\"offel (1999, BE), applied to adaptive optics and HST/STIS
data, indicated that the base of jets are significantly ionized with $x_{\rm e}
\sim 0.01-0.4$ (Bacciotti 2001).  The typical X-ray luminosity from
an active YSO ionizes the x-wind flow at this level.  However, for jets to be
bright in the optical forbidden lines, maintaining $T \sim 5,000 - 10,000$\,K
at thousands of AU from the stars with sufficient emitting areas, poses a
difficulty for the internal heating sources due to powerful adiabatic
expansion.

We also implemented a phenomenological model of wind heating based
on stochastic shock dissipation, similar to MHD simulations of
turbulence (e.g., Ostriker et al. 1999).  The mechanical heating
arises from fluctuations produced by the time dependent processes
that are carried by the wind to large distances where they are
dissipated in shocks, MHD waves, and turbulent cascades.  This
combines the local shock diagnostics that have long been applied
to explain the complex emission spectra (see, e.g., Hartigan,
Bally, Reipurth, and Morse 2000), and the global properties of the
jets that seem closely connected to the activities of the exciting
sources.  Applied to a time-steady flow, the time-averaged effect
of heating can be expressed by the scale-free volumetric heating
rate, $\Gamma_{\rm mech} = \alpha \rho v^3 s^{-1}$, where $\rho$
and $v$ are the local mass density and flow speed, and $s$ is the
distance from the origin.  In particular, $\alpha$ is a
phenomenological constant, whose magnitude characterizes the
strength of the disturbances.  When a partially-revealed but
active YSO is considered, $\alpha \sim 10^{-3}$ in the numerical
calculations produces temperatures and electron fractions that are
high enough for the x-wind jet to radiate in the optical forbidden
lines at the intensity level and on the spatial scales that are
observed (see Figure 1).  Expressed in terms of velocity
variations, such values of $\alpha$ imply a fractional velocity
change of $\lesssim 5\%$, which suggests that the underlying
background steady-state flow is relatively unperturbed.

Figure~2 shows the synthetic image made of the excitation
conditions in Figure~1.  Compared with the synthetic image of
uniform excitation (which only illustrates the integrated column
density out of 3-D flow cube), the appearance of conic shape at
the base of the jet, as has been widely adopted as the ``jet
opening'' in images, is a result of the self-consistent excitation
profile obtained by the phenomenological approach of heating.
This illustrates the importance of self-consistent density and
thermal profiles in the modelling scheme to capture the features
in real images.  The jet width extracted from Figure~2 is very
similar to some jet widths obtained by HST and by adaptive optics
(Dougados et al 2000).

The value of such diagnostics is best illustrated by cross line
ratios from the optical forbidden lines.  Figure 3 is the line
ratio derived from Figure~2. Each blue point is a ``theoretical''
data point from each pixel.  The stars are compiled from the HH
objects available from ground-based data back in Raga et al
(1996).  The circles are DG Tau at resolution approaching
0.1$\arcsec$ for the microjets (read from Fig.3 of
Lavalley-Fouquot et al 2000). The synthetic line ratio plot of one
single jet object, using x-wind model, could encompass the diverse
excitation condition in most objects. The only points that fall
outside of the theoretical domain actually come from strong shock
emissions which cannot be explained in the current theory of weak
shocks. Figure~3 provides the critical test for theoretical models
to generate conditions that excite the observed lines, and opens
up the window for new perspective of interpretation.

\begin {figure}[t]
\epsscale{0.8} \plotone{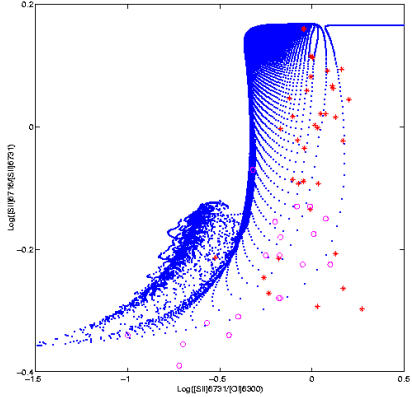} \label{fig3.ps} \caption{[S\,II]
$\lambda$6716/[S\,II] $\lambda$6731 line ratio vs. the [S\,II]
$\lambda$6731/[O\,I] $\lambda$ 6300 based on the synthetic images
in Figure 2, where the jet is viewed perpendicular to its axis. }
\end{figure}

\section{SUMMARY}
In this short article, we summarized recent theoretical
developments that lead to direct comparison with real data by
first-principle modelling of MHD x-winds and physical processes
occurring near the YSOs. We argue that at their base, YSO jets are
optical illusions associated with the excitation mechanisms by
which atomic forbidden lines are excited.  Gentle
time-variabilities or pulses implied by knots in observed jets
mainly contribute to the variation of thermal excitation on top of
background flow, whose densities and velocities are closely
maintained by steady state values. This is a demonstration that,
first-principle calculations of self-consistent physical processes
in young stellar winds provide robust theoretical help on probing
the jet phenomena.

\acknowledgments{}
H.S. sincerely thank the hospitality of the local organizing committee.

\end{document}